\documentclass[nopreprintline,12pt]{elsarticle}

\usepackage{tikz}
\usepackage{float}
\usetikzlibrary{intersections, calc, arrows, automata, positioning, shapes, trees, arrows.meta, fit, shapes.geometric}
\usepackage{bbding}
\usepackage{pifont}
\usepackage{pgf, verbatim}
\usepackage{amssymb}
\usepackage[ruled,vlined,linesnumbered,resetcount]{algorithm2e}
\usepackage{setspace}
\usepackage{wrapfig}
\SetAlgoInsideSkip{smallskip}
\newcommand\Omicron{\mathrm{O}}

\usepackage{amsthm}

\usepackage{lineno}

\usepackage[bookmarks,unicode,colorlinks=true]{hyperref}%
   \def\@citecolor{blue}%
   \def\@urlcolor{blue}%
   \def\@linkcolor{blue}%

\journal{}

\begin{document}

\begin{frontmatter}

\title{Errata to: "Faster Deterministic Exponential
Time Algorithm for Energy Games and Mean Payoff Games"}

\author{Peter Austin, Daniele Dell'Erba}

\address{
            University of Liverpool,
            Liverpool,
            UK
            }

\begin{abstract}
An improved exponential time algorithm for Energy Games and Mean Payoff Games has been recently proposed in ICALP 19. The new algorithm prevents some of the repetitive operations performed by the classic value iteration algorithm of Brim et al., leading to an approach with time complexity $\Omicron(\min(mnW,mn2^{n/2}\log W))$.
Unfortunately, the pseudo-code of the algorithm includes inaccuracies that violate two Lemmata used in the correctness and complexity proofs. In this technical report, we describe the problems, propose a fixed version of the algorithm, and correct  the proofs for the Lemmata.

\end{abstract}


\end{frontmatter}

\newcommand{\LSet}{\mathrm L}
\newcommand{\BSet}{\mathrm B}
\newcommand{\VSet}{\mathrm V}
\newcommand{\nat}{{\mathbb N}}
\newcommand{\UpdEn}{\textsc{Update-Energy}}
\newcommand{\ComEn}{\textsc{Compute-Energy}}
\newcommand{\Upd}{\textsc{Update}}
\newcommand{\Delt}{\textsc{Delta}}

\newtheorem{lm}{Lemma}
\newtheorem*{lma}{Lemma}
\newproof{pf}{Proof}
\newcommand{\Inf}{\mathsf{Inf}}

\section{Introduction}

Energy games~\cite{CAHS03} and mean-payoff games~\cite{EM79} are infinite-duration two-player games on weighted directed graphs. The game starts at an arbitrary vertex from which an infinite path (also called play) is built.
The vertices of the graph are partitioned between the two players: Min and Max. The player controlling a vertex decides how to extend the path by choosing a move, which means, selecting an outgoing edge from the current vertex on the graph.
The payoff of any prefix of a play is the sum of the weights of its edges. The goal of player Max, in an Energy game, is to maximize the payoff of the play such that it never falls below a given initial energy credit.
Player Min, instead, aims at minimising the payoff and possibly let it fall below the initial energy credit.
Whenever the energy constraint applies only to infinite plays, the problem reduces to a mean-payoff game. The two problems are known to be log-space equivalent~\cite{BFLMS08}.
Many algorithms have been devised for solving energy games and mean-payoff games~\cite{LP07,BCDGR11,Sch08a,BDM20a}. In particular, in ICALP 19, Dorfman et al.~\cite{DKZ19} presented a refinement of the value iteration algorithm of Brim et al..~\cite{BCDGR11} in which they put in place two improvements.
First, they identify sequences of update operations that the original approach would perform one by one, and instead, the new approach apply them in a single step.
Second, they apply the scaling technique~\cite{GT91}, which allows to convert the solution recursively obtained on a simpler instance of the game, in which the weights have been halved, to the original game.

Unfortunately, the published version of the paper contains three mistakes that lead to the falsification of two Lemmata and, on cascade, many of the remaining statements do not hold, including the correctness and complexity results.
Fortunately, it is possible to fix these issues. Here, we provide an explanation of each problem with an accompanying counterexample and then propose a correct version of the algorithm. We also provide full proofs for the two Lemmata, and an additional Lemma that is necessary to support the other statements.

\section{Preliminaries}\label{sec:pre}

A weighted game graph is a tuple $G=\langle V_0, V_1, E, w \rangle$, where the set of vertices $V$ is the disjoint union of $V_0$ and $V_1$, $E$ is the edge relation $V\times V$, and $w$ is the weight function $E\rightarrow \mathbb{Z}$. Every vertex has at least one successor and $V_0$ (resp. $V_1$) is the set of vertices controlled by player Max (resp. Min).
Given a function $f:V \rightarrow\mathbb{N}\cup\infty$, the modified weight of an edge $(v,u)$ is denoted by $w_f(v,u)=f(u)-f(v)+w(v,u)$. An edge $(v,u)$ is said to be \emph{valid} w.r.t. $f$ if $w_f(v,u)\geq 0$.
When $w_f(v,u)=0$, the edge is also said to be \emph{tight}.
A vertex $v\in V_0$ (resp. $V_1$) is said to be \emph{valid} w.r.t. $f$ if for some (all) $u\in E(v)$ the edge $(v,u)$ is valid.
A function $f:V \rightarrow\mathbb{N}\cup\infty$, is said to be a solution for $G$ if for every $v\in V$ it holds that: $w_f(v,u)\geq 0$ for some $u\in E(v)$ if $v\in V_0$, and $w_f(v,u)\geq 0$ for all $u\in E(v)$ if $v\in V_1$.

\section{The Algorithm}\label{sec:salg}

Before discussing the problems of the original \textsc{Dorfman-Kaplan-Zwick} (DKZ for short) algorithm, we briefly describe the high-level approach.
The full pseudocode, as presented in~\cite{DKZ19}, is reported in Appendix. In this section, we only report the function that contains the bugs.
The DKZ algorithm computes a solution function $f$ that is a progress measure for player Min.
It follows that for vertices won by player Min, the value returned by $f$ is, in principle, unbounded, while for player Max, it is finite.
To overcome the unboundedness problem, DKZ proposed a preprocessing of the game that ensures all the values remain finite. The preprocessing works in two steps.
First, all negative cycles composed of vertices of player Min are removed from the game. Then, a new sink vertex $s$ with a self loop of weight 0 is added to the game, and all the vertices of player Max are connected to $s$ with an edge of weight $-2nW$, where $n$ is the number of the vertices in the game after the first step and $W$ is the largest absolute weight.

\begin{wrapfigure}[15]{r}{0.755\textwidth}
\vspace{-1.2em}
\begin{algorithm}[H]
	\setstretch{0.95}
	\DontPrintSemicolon
	\NoCaptionOfAlgo
	\caption{\textsc{Update-Energy ($V_0,V_1,E,w,f,v$).}}
	\textbf{if} $v\!\in\! V_0$ $and$ $\forall(v,u)\in E: w_f(v,u) \!<\! 0$ \textbf{then} $L\gets \{v\}\!\!\!\!\!\!\!\!$\;
	\textbf{if} $v\!\in\! V_1$ $and$ $\exists(v,u)\in E: w_f(v,u) \!<\! 0$ \textbf{then} $L\gets \{v\}\!\!\!\!\!\!\!\!$\;
	\ForEach{$u\in V_0$}{
		$count[u]\gets|\{u'$ $|$ $(u,u')\in E,$ $w_f(u,u')\geq 0\}|$
	}
	\While{$L=\{v\}$}{
		$B\gets \{v\}$\;
		\textsc{Update($v,L,B$)}\;
		\While{$L\setminus \{v\}\not=\emptyset$}{
			pick $u\in L\setminus\{v\}$\;
			\textsc{Update($u,L,B$)}
		}
		$\Delta\gets$ \textsc{Delta}($B$)\;
		\textbf{foreach} $u\in B$ \textbf{do} $f(u)\gets f(u)+\Delta$
	}	
\end{algorithm}
\end{wrapfigure}

Once this \emph{finite} game has been constructed, the DKZ algorithm solves it in two phases. In the first one, function \textsc{Compute-Energy} simply applies the scaling technique by recursively solving the game on halved weight. As a base case, the recursion stops when all weights are non-negative due to the rounding up when halving the weights.
It is worth observing that after the recursive call, when both the solution and the weight functions are doubled, all the vertices are valid.
However, the weight function might be an overestimation of the real one due to the rounding. Therefore, the weight of the outgoing edges of a currently focused vertex $v$ are decremented by 1 in case the weight of its edges has been previously rounded.
After this step, the only vertex that can be invalid is $v$. Because of this, \textsc{Update-Energy} is called on $v$ to update the solution function.

The second solving phase is implemented in function \textsc{Update-Energy}. This function, given an input game $G$ in which there is at most one invalid vertex w.r.t. the solution function $f$, returns a valid $f$ on $G$.
This is done by focusing on the invalid vertex $v$ and updating the solution function $f(v)$ so that $v$ is now valid.
The update can make other vertices invalid, therefore, \textsc{Update-Energy} computes the set of vertices $\BSet$ that are \emph{rooted} in $v$.
By rooted we refer to vertices whose value in the solution function currently depends on the one of $v$. In a simple case, it suffice to propagate the update on all the vertices in $\BSet$ to obtain a valid $f$. Otherwise, multiple updates are required.
In this more complex case, instead of iterating several times over $\BSet$, the number of updates $\Delta$ that is required to make the vertices valid is predicted and applied in a single update operation to $\BSet$. The computation of $\Delta$ is done by means of the function $\Delt$. In particular, $\Delta$ corresponds to the minimum of three different parameters that together ensure that after the $\Delta$-lift, that is the update of value $\Delta$ of the solution function $f$, no new vertex became invalid, and those who are valid are also tight.
To make the implementation more efficient, during \textsc{Update-Energy}, a vector called $count$ keeps track of the number of tight edges for every vertex of player Max.

\section{The Errors}\label{sec:err}

In this section we describe the three errors of the algorithm that induce violations of two Lemmas. We also provide games that exploit the errors to show that the bugs are realizable and that they lead to wrong solutions or different complexity bounds.
In order to show the violations, we report the two problematic Lemmas. The first one states that, during the execution of \textsc{Update-Energy}, $\LSet$ contains all and only invalid vertices.
\begin{lm}[3.3 \cite{DKZ19}]
\label{thm:invalid}
During $\UpdEn$, $u \in \LSet$ if and only if $u$ is invalid.
\end{lm}

The second Lemma states that whenever a vertex $v$ has been updated, i.e. in the solution function $f$ it has a positive value, then if $v$ does not belong to $\LSet$, it must be tight.
\begin{lm}[3.4 \cite{DKZ19}]
\label{thm:tight}
During $\UpdEn$, if $u \not\in \LSet$ and $f(u) > 0$, then $u$ is tight.
\end{lm}
Note that by Lemma~\ref{thm:invalid}, the vertices not in $\LSet$ are valid. Hence, the combination of the two Lemmas implies that $f$ is the least solution function.

\subsection{Delta Call}

The first issue is related to the use of function $\Delt$.
Let us consider a game $G$ having an invalid vertex $v$ that is detected and analysed by \textsc{Update-Energy}.
Due to the use of the scaling technique, it is sufficient to increment $f(v)$ by $1$ to make $v$ valid, via a single update. 
This observation follows from the fact that the game has been already solved with halved weights and only the adjustment to weights, at Line 8 of \textsc{Compute-Energy}, can make a vertex invalid for a gap of $-1$. Hence, after \textsc{Update} is called at Line 7 of \textsc{Update-Energy}, $v$ is valid. However, other vertices can now be invalid. Therefore, they are updated in the while loop at Lines 8-10.
The condition for the loop to end is that all the vertices, except $v$, are valid.
It follows that when $\Delt$ is called the only vertex that is eventually invalid is $v$. In both cases $\Delt$ is called.
We claim that $\Delt$ must be called if and only if $v$ is invalid.
Otherwise, it is possible for $\Delt$ to compute a positive $\Delta$ that makes $v$ not tight, violating Lemma~\ref{thm:tight}.

Unfortunately, this problem is realisable and to show it we have built the game depicted in Figure~\ref{fig:errd}. On the left, we show the game with the original weights, while on the right, is shown the game analyzed by \textsc{Update-Energy} when the problem occurs.
In the below notation, diamonds vertices belong to player Max (who wants to keep the values of the solution $f$ between 0 and $nW$), while squares vertices belong to player Min (that, instead, wants to reach higher values than $nW$).

\begin{figure}[hbt]
  \centering
  \scalebox{0.85}{
    \begin{tikzpicture}[-{Stealth[length=3mm, width=2mm]},node distance={30mm}, thick, mainsq/.style = {draw, minimum size = 1cm}, maindiam/.style = {draw, diamond, minimum size = 1.2cm}]
        \node[mainsq] (node1) at (0,0){\Large$v$};
        \node[maindiam] (node0) at (3,0){\Large$u$};
        \node[maindiam] (sink1) at (6,0){\Large$s$};
        
        \path[every loop/.append style=-{Stealth[length=3mm, width=2mm]}]
        (node0) edge node[above]{\Large$-4$} (sink1)
        (node0) edge [bend left=20] node[below]{\Large$1$} (node1)
        (node1) edge [bend left=20] node[above]{\Large$-1$} (node0)
        (sink1) edge [loop below] node[below]{\Large$0$} (sink1)
        ;
    \end{tikzpicture}
    }
  \hspace{1em}
  \scalebox{0.85}{
    \begin{tikzpicture}[-{Stealth[length=3mm, width=2mm]},node distance={30mm}, thick, mainsq/.style = {draw, minimum size = 1cm}, maindiam/.style = {draw, diamond, minimum size = 1.2cm}]
        \node[mainsq] (node1) at (0,0){\Large$v$};
        \node[maindiam] (node0) at (3,0){\Large$u$};
        \node[maindiam] (sink1) at (6,0){\Large$s$};
        \node[draw=red, style=dashed, fit=(node1), inner sep=4mm](Fit1) {};
        
        \path[every loop/.append style=-{Stealth[length=3mm, width=2mm]}]
        (node0) edge node[above]{\Large$-4$} (sink1)
        (node0) edge [bend left=20] node[below]{\Large$2$} (node1)
        (node1) edge [bend left=20] node[above]{\Large$-1$} (node0)
        (sink1) edge [loop below] node[below]{\Large$0$} (sink1)
        ;
    \end{tikzpicture}
    }
	\caption[Game 1]{On the left, a problematic game for $\Delt$ check; on the right, the snapshot of the game when $\UpdEn$ is called on $v$ and $f = \langle 0,0,0 \rangle$}
 \label{fig:errd}
\end{figure}

\textbf{Example 1.} When the last recursive \textsc{Compute-Energy} call ends, on the input game shown at the left side of Figure~\ref{fig:errd}, it returns the solution $f = \langle 0,0,0 \rangle$ for the game with halved weights: $w(v,u)=0; w(u,v)=1; w(u,s)=-2; w(s,s)=0$. Then, at Line 5 of \textsc{Compute-Energy}, both $f$ and $w$ are doubled, and the foreach loop picks the vertex $v$, scaling back of $-1$ the weight of the edge $(v,u)$. The resulting game is depicted on the right side of Figure~\ref{fig:errd}.
At this point, \textsc{Update-Energy} is called on $v$ that is invalid since $w_f(v,u)=f(v)-f(u)+w(v,u)=0-0-1=-1$. As a consequence, $v$ is firstly added to $\LSet$ and then updated by $\Upd$, lifting the solution function to $\langle 1,0,0 \rangle$.
The update does not makes $u$ invalid, hence, at the end of the while loop we have that $\BSet=\{v\}$ (as highlighted by the red box in the picture).
When $\Delt$ is called at Line 11, it computes the values: $p_1=\infty$, since $\BSet\cap V_0$ is empty; $p_2=1$, since the edge $(u,s)$ enable the computation of $\gamma(u)$ that contributes with value $w_f(u,v)=0+1-2=-1$; and $p_3=\infty$, since $\bar\BSet\cap V_1$ is empty.
Thus, $\Delta=1$, and the solution function is lifted again for $v$, leading to the non tight solution $\langle 2,0,0 \rangle$. It is worth noting that in principle this \emph{overflow} can wrongly assign $v$ as winning for the opponent player. However, if we keep the algorithm running, at the next iteration, it will compute a negative $\Delta$, fixing the tightness problem. Unfortunately, this type of operation breaks the complexity bounds, allowing the solution function to be not increasing.

\subsection{Counters update}

A second issue is due to a wrong use of the vector $count$. We recall that $count$ keeps track of the number of tight edges of every vertex of player Max. This means that whenever a vertex is updated, also $count$ should be modified accordingly. In the algorithm there are two sources of update: one in function $\Upd$, and one as a result of the $\Delta$-lift when $\Delt$ is called. In the first case, at Line 3 of $\Upd$, $count$ is updated, while in the second case it is not. This might lead, in the next iteration of the while loop at Lines 5-12, to 
wrongly trigger the update of vertices that are valid. This would violate Lemma~\ref{thm:invalid}.

\begin{figure}[hbt]
  \centering
  \scalebox{0.85}{
    \begin{tikzpicture}[-{Stealth[length=3mm, width=2mm]},node distance={30mm}, thick, mainsq/.style = {draw, minimum size = 1cm}, maindiam/.style = {draw, diamond, minimum size = 1.2cm}]
        \node[maindiam] (node0) at (3,0){\Large$u$};
        \node[mainsq] (node1) at (0,0){\Large$v$};
        \node[maindiam] (node2) at (3,-3){\Large$w$};
        \node[maindiam] (sink1) at (6,0){\Large$s$};
        
        \path[every loop/.append style=-{Stealth[length=3mm, width=2mm]}]
        (node0) edge [bend left=20] node[below]{\Large$0$} (node1)
        (node0) edge node[right]{\Large$-2$} (node2)
        (node0) edge node[above]{\Large$-16$} (sink1)
        (node1) edge [bend left=20] node[above]{\Large$-1$} (node0)
        (node2) edge node[below right]{\Large$-16$} (sink1)
        (node2) edge [loop left] node[left]{\Large$1$} (node2)
        (sink1) edge [loop below] node[below]{\Large$0$} (sink1)
        ;
	\end{tikzpicture}
    }
  \hspace{1em}
  \scalebox{0.85}{
        \begin{tikzpicture}[-{Stealth[length=3mm, width=2mm]},node distance={30mm}, thick, mainsq/.style = {draw, minimum size = 1cm}, maindiam/.style = {draw, diamond, minimum size = 1.2cm}]
            \node[maindiam] (node0) at (3,0){\Large$u$};
            \node[mainsq] (node1) at (0,0){\Large$v$};
            \node[maindiam] (node2) at (3,-3){\Large$w$};
            \node[maindiam] (sink1) at (6,0){\Large$s$};
            \node[draw=red, style=dashed, fit=(node1)(node0), inner sep=4mm](Fit1) {};
            
            \path[every loop/.append style=-{Stealth[length=3mm, width=2mm]}]
            (node0) edge [bend left=20] node[below]{\Large$0$} (node1)
            (node0) edge node[right]{\Large$-2$} (node2)
            (node0) edge node[above]{\Large$-16$} (sink1)
            (node1) edge [bend left=20] node[above]{\Large$-1$} (node0)
            (node2) edge node[below right]{\Large$-16$} (sink1)
            (node2) edge [loop left] node[left]{\Large$1$} (node2)
            (sink1) edge [loop below] node[below]{\Large$0$} (sink1)
            ;
        \end{tikzpicture}
    }
	\caption[Game 1]{On the left, problematic game for $count$ check; on the right, snapshot of the game when $\UpdEn$ is called on $v$ and $f = \langle 0,0,0,0 \rangle$}
 \label{fig:errc}
\end{figure}
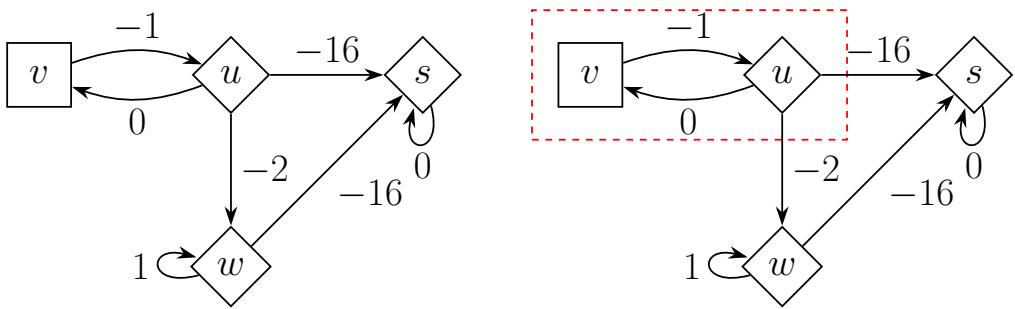

To realise this problem we add one vertex to the example depicted above, as shown in Figure~\ref{fig:errc}. In this case, the error occurs when the weights of the game have their real value, i.e. the weights are not halved.

\textbf{Example 2.} The $counter$ vector problem occurs, on the game depicted in Figure~\ref{fig:errc}, after the last recursive call of \textsc{Compute-Energy}, when all the weights have been restored to their original value.
Once \textsc{Update-Energy} is called on $v$, that is invalid since $w_f(v,u)=-1$, $f(v)$ is updated making $f= \langle 1,0,0,0 \rangle$ and $count=\langle 1,0,1,1 \rangle$.
This update makes $u$ invalid because now all its outgoing edges have negative modified weight, indeed, $count(u)=0$. The minimum value among them is $w_f(u,v)=-1$. Hence, $u$ is firstly added to both $\LSet$ and $\BSet$, and then updated making $f= \langle 1,1,0,0 \rangle$ and $count=\langle 0,1,1,1 \rangle$. Therefore, $v$ is again invalid and the algorithm proceeds by calling function $\Delt$ on $\BSet=\{u,v\}$, as highlighted by the red box of Figure~\ref{fig:errc}.
The three parameters of $\Delta$ corresponds to: $p_1=1$, since $w_f(u,w)=-1$; $p_2=\infty$, since there are no vertices in $\bar\BSet\cap V_0$ that satisfy the condition; and $p_3=\infty$, since there are no edges from $\bar\BSet\cap V_1$ to $\BSet$. Thus, $\Delta=1$ and $\BSet$ is lifted making $f= \langle 2,2,0,0 \rangle$. Note that now $u$ is valid and its counter should be set to $1$, but this update of the vector is missing.
As a consequence, since $v$ is still invalid, another iteration of the while loop begins. At Line 7, $v$ gets updated, making $f= \langle 3,2,0,0 \rangle$ and $count=\langle 1,0,1,1 \rangle$.
Then, at Line 10, $u$ is wrongly updated again, making $f= \langle 3,3,0,0 \rangle$, despite $w_f(u,w)=0$.

\subsection{Set of invalid vertices}

A final issue arises when performing a $\Delta$-lift.
Due to the check at Line 11 and the while condition at Line 8, we know that before the $\Delta$-lift, $v$ is the only invalid vertex.
Now two cases may arise. If the value computed by $\Delt$ is zero, then $v$ is still invalid. Otherwise, if $\Delta$ is positive, then the $\Delta$-lift can make $v$ valid. In the latter case, $v$ still belongs to $\LSet$ and then in the next iteration of the loop it gets wrongly updated, violating Lemma~\ref{thm:tight} for which valid vertices must be tight.
To be precise, if the $\Delta$-lift makes $v$ valid, then Lemma~\ref{thm:invalid} does not holds since $v$ is kept in $\LSet$. This observation immediately leads to the fix that will be applied in Section~\ref{sec:fve}.

\begin{figure}[hbt]
  \centering
  \scalebox{0.85}{
    \begin{tikzpicture}[-{Stealth[length=3mm, width=2mm]},node distance={30mm}, thick, mainsq/.style = {draw, minimum size = 1cm}, maindiam/.style = {draw, diamond, minimum size = 1.2cm}]
        \node[mainsq] (node1) at (0,0){\Large$v$};
        \node[maindiam] (node0) at (3,0){\Large$u$};
        \node[maindiam] (sink1) at (6,0){\Large$s$};
        
        \path[every loop/.append style=-{Stealth[length=3mm, width=2mm]}]
        (node0) edge node[above]{\Large$-4$} (sink1)
        (node0) edge [bend left=20] node[below]{\Large$-1$} (node1)
        (node1) edge [bend left=20] node[above]{\Large$-1$} (node0)
        (sink1) edge [loop below] node[below]{\Large$0$} (sink1)
        ;
    \end{tikzpicture}
    }
  \hspace{1em}
  \scalebox{0.85}{
    \begin{tikzpicture}[-{Stealth[length=3mm, width=2mm]},node distance={30mm}, thick, mainsq/.style = {draw, minimum size = 1cm}, maindiam/.style = {draw, diamond, minimum size = 1.2cm}]
        \node[mainsq] (node1) at (0,0){\Large$v$};
        \node[maindiam] (node0) at (3,0){\Large$u$};
        \node[maindiam] (sink1) at (6,0){\Large$s$};
        \node[draw=red, style=dashed, fit=(node1)(node0), inner sep=4mm](Fit1) {};
        
        \path[every loop/.append style=-{Stealth[length=3mm, width=2mm]}]
        (node0) edge node[above]{\Large$-4$} (sink1)
        (node0) edge [bend left=20] node[below]{\Large$-1$} (node1)
        (node1) edge [bend left=20] node[above]{\Large$0$} (node0)
        (sink1) edge [loop below] node[below]{\Large$0$} (sink1)
        ;
    \end{tikzpicture}
    }
	\caption[Game 1]{On the left, problematic game for the validity invariant of $\LSet$; on the right, snapshot of the game when $\UpdEn$ is called on $u$ and $f = \langle 0,0,0 \rangle$}
 \label{fig:errl}
\end{figure}
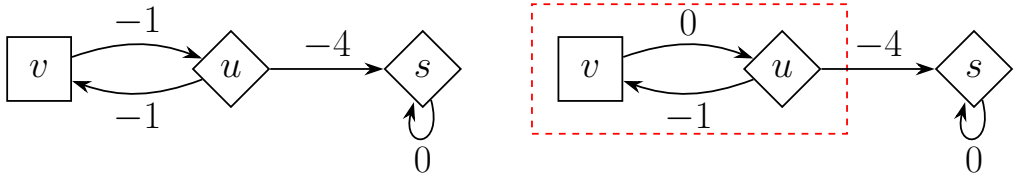

To show this problem, we have modified the game used for the first error, by changing the sign of the weight of the edge $(u,v)$, as shown in Figure~\ref{fig:errl}.

\textbf{Example 3.} When executing the algorithm on the game  depicted in Figure~\ref{fig:errl}, we focus on the last recursive \textsc{Compute-Energy} call, that ends with $f=\langle 0,0,0 \rangle$ and the following weights: $w(v,u)=0; w(u,v)=0; w(u,s)=-2; w(s,s)=0$. At Line 5 of \textsc{Compute-Energy}, $w$ is doubled, and the for each loop picks $u$ that is invalid since $w_f(u,v)=-1$ and $w_f(u,s)=-4$. At Line 7, $u$ is updated making $f=\langle 0,1,0 \rangle$, while $v$ become invalid because now $w_f(v,u)=-1$. At Line 10, $u$ is also updated, making $f=\langle 1,1,0 \rangle$ and $u$ invalid again.
At this point $\BSet=\{u,v\}$, as highlighted by the red box of Figure~\ref{fig:errl}, and $\Delt$ is called.
The three parameters of $\Delta$ are: $p_1=3$, since $w_f(u,s)=-3$; $p_2=\infty$, since there are no vertices in $\bar\BSet\cap V_0$ that satisfy the condition; and $p_3=\infty$, since 
$\bar\BSet\cap V_1$ is empty. Thus, $\Delta=3$ and $\BSet$ is lifted making $f= \langle 4,4,0 \rangle$. Observe that $u$ is valid and, by Lemma~\ref{thm:invalid}, should be removed from $\LSet$. Due to this missing check, another iteration of the while loop begins, that will wrongly updates $u$, violating also Lemma~\ref{thm:tight}.

\section{Fixed version}\label{sec:fve}

We finally present a correct version of the algorithm along with proofs for the two Lemmas, and introduce a new Lemma to support the other statements.

\begin{wrapfigure}[18]{r}{0.755\textwidth}
\vspace{-1.2em}
\begin{algorithm}[H]
	\setstretch{0.95}
	\NoCaptionOfAlgo
    \DontPrintSemicolon
	\caption{\textsc{Update-Energy ($V_0,V_1,E,w,f,v$).}}
	\lIf{$v\!\in\! \VSet_0$ $and$ $\forall(v,u)\in E: w_f(v,u) \!<\! 0$} {
    $\!\!\LSet\gets \{v\}\!\!\!\!\!\!\!\!\!\!$
    }
	\lIf{$v\!\in\! \VSet_1$ $and$ $\exists(v,u)\in E: w_f(v,u) \!<\! 0$} {
    $\!\!\LSet\gets \{v\}\!\!\!\!\!\!\!\!\!\!$
    }
	\While{$\LSet=\{v\}$}{
        \ForEach{$u\in \VSet_0$}{
            $count[u]\gets|\{u'$ $|$ $(u,u')\in E,$ $w_f(u,u')\geq 0\}|$
        }
		$\BSet\gets \{v\}$\;
		\textsc{Update($v,\LSet,\BSet$)}\;
		\While{$\LSet\setminus \{v\}\not=\emptyset$}{
			pick $u\in \LSet\setminus\{v\}$\;
			\textsc{Update($u,\LSet,\BSet$)}
		}
        \If{$v\in \LSet$}{
    		$\Delta\gets$ \textsc{Delta}($\BSet$)\;
            \lForEach {$u\in \BSet$} {$f(u)\gets f(u)+\Delta$}
            \lIf{$v$ is valid}{
    		$\LSet\gets\LSet\setminus\{v\}$
            }
        }
	}
 \label{alg:fix}
\end{algorithm}
\end{wrapfigure}

All the fixes have been applied to \textsc{Update-Energy}, whose pseudocode is reported in Algorithm~\ref{alg:fix}.
The first error has been fixed by checking whether $v$ is valid before calling $\Delt$, at Line 11. The issue related to the $count$ vector has been solved by moving the update of the vector inside the while loop, at Line 4-5.
In this way, $count$ is updated before the call to $\Upd$, ensuring that the values stored in it are consistent with the number of valid edges.
Also the last error has a trivial fix, that obviously follows from the observations made in Section~\ref{sec:err}. To guarantee Lemma~\ref{thm:invalid} holds after the application of a $\Delta$-lift, it is necessary to check whether $v$ is still invalid at Line 14. Otherwise, it has to be removed from $\LSet$. 

To prove that the properties stated in the two Lemmas hold in the fixed version of \textsc{Update-Energy}, it is necessary to prove that they hold after the $\Delta$-lift of the set of vertices $\BSet$. This argument is missing in the original paper. To make the new proofs 
complete and readable, we separately prove the properties for the $\Delta$-lift in a new Lemma.

\begin{lm}[Delta Guarantees]
\label{thm:delta}
In $\UpdEn$, if $\Delt$ is called, then after the $\Delta$-lift of the vertices in $\BSet$: (1) no vertex becomes invalid, and (2) all valid vertices with a positive value in $f$ are tight. 
\end{lm}
\begin{pf}
To show that after the $\Delta$-lift of
$\BSet$ there are no new invalid vertices, we recall that
$v$ is the only invalid vertex before the lift.
The vertices in $\BSet$, instead, have been lifted by $\Upd$
and made valid. Hence, every $u\in\BSet$ has at least one tight edge $(u,u')$ with $u'\in\BSet$. As a consequence, if both $u$ and $u'$ are lifted for the same value $\Delta$, then
$(u,u')$ will still be tight, and
$u$ still valid. 
To avoid that the $\Delta$-lift makes vertices in $\overline{\BSet}$ invalid, 
the value of $\Delta$ is bounded by $p_3$, that corresponds to the gap between the solution $f$ of any $u\in\VSet_1\cap\overline{\BSet}$ that has an edge $(u,u')$ toward $\BSet$ and the solution of a successor $u'\in\BSet$. Such a check does not involve vertices in $u\in\VSet_0\cap\overline{\BSet}$ with positive $f$, as none of these vertices can have a single and only tight edge toward $\BSet$, and therefore, they will keep at least one tight edge after the lift of the vertices in $\BSet$.
For contradiction let us assume there is a $u\in\VSet_0\cap\overline{\BSet}$ with a positive $f$ and a single tight edge toward
$u'\in\BSet$. Observe that all the vertices in $\BSet$ have been lifted by $\Upd$.
Hence, before the lift, the edge $(u,u')$ was valid but not tight, which is a condition possible only for the vertices that have never been lifted, namely those having $f(v)=0$.
In fact, vertices $v$ in $\VSet_0\cap\overline{\BSet}$ with $f(v)=0$, can be valid and not tight if there is a successor $u'$ with positive modified weight and, therefore, a valid edge. In this latter case, $u'$ belongs to $\BSet$ and, therefore, the $\Delta$-lift of $\BSet$ can make the edge invalid.
For this reason, $\Delta$ is bounded also by $p_2$.
This second component of $\Delta$ focuses on the vertices of player Max with no valid edges toward $\overline{\BSet}$. Since these vertices are valid, they must have a valid edge toward $\BSet$ that is not necessarily tight.
It follows that
$\BSet$ can be lifted by a value that is bounded by the gap of the modified weight of the outgoing edges from the vertices in $\VSet_0\cap\overline{\BSet}$.

The second condition of the lemma requires that after the $\Delta$-lift of 
$\BSet$, all valid vertices with positive $f$ are tight. Let us first focus on $\overline{\BSet}$.
For the vertices of player Min, as observed in the previous part of the proof, $p_3$ bounds the value of $\Delta$ so that any vertex $u\in\VSet_1\cap\overline{\BSet}$ remains valid. Since these vertices are not lifted by $\Delt$, they remain also tight.
The only vertices of player Max with a valid edge toward $\BSet$ can be those with $f$ equal to 0, as observed in the first part of the proof.
The gap between these positions and the one in $\BSet$ corresponds to 
$p_2$.
We now turn our attention to 
$\BSet$. For 
player Min, it holds that the vertices have at least one tight edge inside $\BSet$. 
Therefore, they will keep such edge tight after the $\Delta$-lift. For 
player Max, instead, $\Delta$
has to be bounded by the gap of the modified weights of the outward edges. This value corresponds to
$p_1$.

\end{pf}

\begin{lma}[3.3 \cite{DKZ19}]
During $\UpdEn$, $u \in \LSet$ if and only if $u$ is invalid.
\end{lma}
\begin{pf}
$(\Rightarrow)$
We need to show that, during $\UpdEn$, $\LSet$ contains only invalid vertices.
We focus on the instructions that add vertices to $\LSet$ or lift the solution $f(u)$ of vertices $u\in\LSet$, making them potentially valid.
At the beginning of the function, at Lines 1-2, a validity check is ran on $v$.
In case $v$ is invalid, then $\LSet=\{v\}$, as required by the lemma. Otherwise the function ends.
At Lines 7 and 10, the function $\Upd$ is called. In the first lines this function increases 
$f(u)$, removes it from $\LSet$, and updates $count$ to ensure that the vector is consistent with the number of valid edges of $u$.
The correctness of these operations follows from the application f the scaling technique. Indeed, before the for each loop of $\ComEn$, all the vertices are valid because the game has been solved by the recursive call and both the weights $w$ and the solution $f$ have been doubled.
As a consequence of the re-scaling at Line 8 however, $v$ might be invalid due to a modified weight equal to -1. Therefore it suffices to increment $f(v)$ to make it valid again. On cascade, any other vertex that is made invalid as a consequence of this increment, can be made valid by means of a single increment. This is done in $\Upd$, where in case the lift of $u$ made some of its predecessors invalid, they are added to $\LSet$, in the for each loop at Lines 5-9. Hence, after the call of $\Upd$, $\LSet$ contains only invalid vertices.
At the end of the while loop of $\UpdEn$ at Lines 8-10, the only vertex that $\LSet$ can contain is $v$. Therefore, at Line 13, $v$ is the only vertex in $\LSet$ that can become valid after the $\Delta$-lift, at Line 13.
After that, the check at Line 14 ensures that in this case $u$ is removed from $\LSet$.
Finally, since the $\Delta$-lift of
$\BSet$ makes the counters outdated, at the beginning of the next iteration of the while loop (Lines 4-5) $count$ is computed again for all the vertices.

$(\Leftarrow)$
We need to show that, during $\UpdEn$, every invalid vertex is in $\LSet$. We focus on the instructions that remove vertices from $\LSet$, or lift the solution $f$, operation that may make other vertices invalid.
We remark that all the vertices but the one selected in the Line 6 of $\ComEn$ are valid.
Hence at the beginning of $\UpdEn$, a validity check is ran on $v$.
When 
$\Upd$ is called on a vertex $u$, at Lines 7 or 10, $u$ is made valid with a single lift and then removed from $\LSet$ (Lines 1-2 of $\Upd$). Moreover, if this lift made other vertices invalid, these are identified and added to $\LSet$ in the for each loop at Lines 4-8.
At Line 12 of $\UpdEn$,
$\Delt$ is called only if $v$, that is the only vertex in $\LSet$, is invalid.
Due to Lemma~\ref{thm:delta}, 
after the $\Delta$-lift of $\BSet$,
no vertex becomes invalid. The vertex $v$, instead, can now valid and consequentially a check is ran to remove it from $\LSet$, at Line 14.
\end{pf}

\begin{lma}[3.4 \cite{DKZ19}]

During $\UpdEn$, if $u \not\in \LSet$ and $f(u) > 0$, then $u$ is tight.
\end{lma}
\begin{pf}
We need to show that during $\UpdEn$, for every vertex $u$ for which $f(u) > 0$, if $u$ is valid then it is tight. 
We focus on the instructions that remove vertices from $\LSet$ or lift the solution
function $f$,
making vertices, potentially, non tight.
Both of these operations occur in function $\Upd$. By Lemma~\ref{thm:invalid}, it holds that every vertex in $\LSet$ is invalid. Hence, after a single increment it cannot become non tight.
When a vertex $u$ is lifted at Line 2, it is tight and can be removed from $\LSet$ (Line 1).
Moreover, in case the increment of
$f(u)$ makes other vertices invalid and, therefore, non tight, they are added to $\LSet$ (Lines 5-9 of $\Upd$).
After the while loop of Lines 8-10 in $\UpdEn$, if $v$ is invalid then
$\Delt$ is called and $\BSet$ is lifted by 
$\Delta$.
(Lines 11-13).
Lemma~\ref{thm:delta} ensures that in this case, all vertices but $v$ that do not belong to $\LSet$ remain valid and are also tight.
A special case applies for the vertices of player Max that have never been updated, e.g. those for which $f(u) = 0$.
These vertices can be valid and not tight if there is a successor $u'$ with positive modified weight and, therefore, a valid edge.
This successor prevents the vertex to lose all its valid edges, even if they are not tight.
\end{pf}

%\section{Conclusions}

\bibliographystyle{elsarticle-num} 
\bibliography{bib}

\newpage
\appendix
\section*{Appendix: DKZ original algorithm}

\begin{algorithm}[H]
	\setstretch{1.1}
	\DontPrintSemicolon
	\NoCaptionOfAlgo
	\caption{\textsc{Compute-Energy ($V_0,V_1,E,w$).}}
	\If{$w\geq 0$}{
		\Return $f\equiv 0$
	}
	$w'\gets \lceil\frac{w}{2}\rceil$\;
	$f\gets$ \textsc{Compute-Energy($V_0,V_1,E,w'$)}\;
	$f,w'\gets 2f,2w'$\;
	\ForEach{$v\in V$}{
		\ForEach{$e\in out(v)$}{
			\textbf{if} $w'(e) > w(e)$ \textbf{then} $w'(e)\gets w'(e)-1$\;
        }
		\textsc{Update-Energy($V_0,V_1,E,w',f,v$)}
	}
	\Return $f$
\end{algorithm}
\begin{algorithm}[H]
	\setstretch{1.1}
    \DontPrintSemicolon
	\NoCaptionOfAlgo
	\caption{\textsc{Update($u,\LSet,\BSet$).}}
    $\LSet\gets \LSet\setminus\{u\}$\;
	$f(u)\gets f(u)+1$\;
	\lIf {$u\in \VSet_0$} {$count[u]\gets |\{(u,u')\in E$ $|$ $w_f(u,u')\geq 0\}|$}
	\ForEach{$u'\in in(u)$ $\textbf{such that}$ $w_f(u',u) < 0$}{
		\If{$u'\in \VSet_0$}{
			\lIf {$w_f(u',u)=-1$} {$count[u']\gets count[u']-1$}
			\lIf {$count[u']=0$} {$\LSet\gets \LSet\cup\{u'\}, \BSet\gets \BSet\cup\{u'\}$}
		}
		\lIf {$u'\in \VSet_1$} {$\LSet\gets \LSet\cup\{u'\}, \BSet\gets \BSet\cup\{u'\}$}
	}	
\end{algorithm}

\begin{algorithm}[H]
	\setstretch{1.1}
	\DontPrintSemicolon
	\NoCaptionOfAlgo
	\caption{\textsc{Delta($B$).}}
	$p_1\gets$ min $\{-w_f(u,u')$ $|$ $(u,u')\in E(B\cap V_0, \bar B)\}$\;
	$p_2\gets$ min $\{\gamma(u)$ $|$ $u\in\bar B\cap V_0, \forall u'\in \bar B$ $w_f(u,u') < 0\}$\;
	$p_3\gets$ min $\{w_f(u,u')$ $|$ $(u,u')\in E(\bar B\cap V_1, B)\}$\;
	\Return{\textup{min} $\{p_1,p_2,p_3\}$}
\end{algorithm}

\end{document}